\documentclass[twocolumn,apj]{emulateapj}
\usepackage{graphicx,natbib}
\shorttitle{orientation of nuclear obscurer}
\shortauthors{Shen, Shao \& Gu}

\begin{document}

\title{The orientation of the nuclear obscurer of the AGNs}
\author{Shiyin Shen$^{1,2}$, Zhengyi Shao$^{1,2}$ \& Minfeng Gu$^{1,3}$}

 \affil{$^1$ Key Laboratory for Research in Galaxies and Cosmology, Shanghai Astronomical
Observatory, Chinese Academy of Sciences, 80 Nandan Road, Shanghai, 200030,
China}
 \email{ssy@shao.ac.cn}
 \affil{$^2$ Key Lab for Astrophysics, Shanghai 200234}
 \affil{$^3$ Department of Physics, University of California, Santa Barbara, CA 93106, USA}

\begin{abstract}
We examine the distribution of axis ratios of a large sample of disk galaxies hosting type 2 AGNs selected from the Sloan Digital Sky Survey
and compare it with a well-defined control sample of non-active galaxies. We find them significantly different, where the type 2 AGNs show both an excess of edge-on objects and deficit of round objects. This systematical bias can not be explained by a nuclear obscurer oriented randomly with respect to the stellar disk. However, a nuclear obscurer coplanar with the stellar disk also does not fit the data very well. By assuming that the nuclear obscurer having an opening angle of $\sim60$ degree, we find  the observed axis ratio distribution can be nicely reproduced by a mean tilt angle of  $\sim30$ degree between the nuclear obscurer and the stellar disk.
\end{abstract}


\maketitle

\section{Introduction}
In the unified AGN model, a type 2 AGN is seen from an angle close to the plane of the central accretion disk. On this plane, the accretion disk and  the broad line regions are obscured by an outer molecular torus, while a bi-polar jet emanating from the nucleus is perpendicular to the plane\citep{Antonucci93}.
The type 1 AGNs are then these objects that we are able to see directly into the central regions.

Several studies have revealed that the angle between the orientation of the kiloparsec-scale jet  and the normal to the host galaxy plane has a wide/random distribution based on the samples of dozens of radio galaxies\citep{Schmitt97,Kinney00,Gallimore06}. These results lead to a conclusion that the  internal gas fuelling the black hole may not be co-aligned with the outer gas forming the stellar disk.

On the other hand, the AGNs are known to be biased sample of inclinations of stellar disks for a long time. Both the optically selected (mostly type 1)  and the soft X-ray selected Seyfert galaxies show a bias against the edge-on galaxies, while the hard X-ray selected sample do not show such a bias\citep{Keel80,Lawrence82,Simcoe97,Zhang09}. \cite{Mcleod95} found an excess of face-on galaxies hosting both type 1 and type 2 AGNs and  explained it with a substantial amount of obscuring material coplanar with the stellar disk\citep[see also][]{Zotti85}.  \cite{Maiolino95} showed that the intermediate type AGNs(1.8,1.9) are mostly found in edge-on galaxies and concluded a 100 pc-scale torus coplanar with the stellar disk.

In this study, we use a large  type 2 AGN sample selected from the Sloan Digital Sky Survey \citep[SDSS]{York00} to check whether the stellar disks of their hosts are biased to edge-on views. We then further check whether this bias is better explained by a dust layer coplanar with the galaxy plane or by an intrinsic alignment between the nuclear obscurer and stellar disk.  This paper is organized as follows. In section 2, we study the axis ratio distribution of the AGNs and make comparisons to that of a well-defined sample of non-active disk galaxies. With the distributions of the shape paraments of the spiral disks concluded in Section 3, we then build toy models to quantify the biased axis ratios of the AGN hosts in Section 4.  Finally, we  make brief conclusions and discussions in Section 5.

\section{the inclinations of the stellar disks of the type 2 AGN hosts}

Our type 2 AGNs are selected from the complete spectroscopic main galaxy sample of the data release 7\citep[DR7,][]{DR7} of SDSS using the empirical relation between the emission line ratios proposed by \citet{Kauffmann03}(see their equation 1). The emission line data are take from the MPA-JHU DR7 release of spectrum measurements(\url{http://www.mpa-garching.mpg.de/SDSS/DR7/}).We use the SDSS parameter `$fracDeV$' and  criteria $fracDev < 0.5$ \footnote{$fracDeV$ is defined in \citet{DR2} through Fcomposite = $fracDeV$ FdeV + (1 - $fracDeV$) Fexp, where Fcomposite, FdeV and Fexp are the composite, de Vaucouleurs and exponential fluxes of the object respectively.}, which ensures that the galaxy flux is dominated  by an exponential(disk) component, to select the AGNs  hosted by disk galaxies. The number of the type 2 AGNs hosted by disk galaxies  is then 32,618.

The inclination of a stellar disk is informed by its apparent axis ratio $b/a$. If a disk is circular and thin, then $b/a=\rm{cos}\, i$, where $i$ is the inclination of the disk (the angle between the line-of-sight and the normal of the disk). If this disk is viewed from arbitrary directions,  the resulted $b/a$ follows a uniform distribution between 0 and 1.

For a more realistic triaxial ellipsoid model, the observed $b/a$ is further dependent on two shape parameters, the disk height $\gamma\equiv C/A$ and the disk ellipticity $\epsilon\equiv(1-B/A)$, where $A,B,C$ are  the major, middle and minor axis of the ellipsoid respectively\citep{Binney85}. The $b/a$ distribution is then dependent on both of  the shape parameter($\gamma, \epsilon$) and viewing angle distributions. The galaxies with different physical properties (e.g. mass and size) are shown to have different intrinsic  shape parameters\citep{Padilla08}. In addition, the sample selection effects could also introduce biases into the measured $b/a$ distribution of a galaxy sample.  For example, the seeing  would make the galaxies with small apparent sizes look rounder\citep{Shao07}. Therefore, to quantify the inclinations of the disks of the AGN hosts, their shape parameters must be well-defined or controlled.

The AGN host galaxies are known to have distinctive physical properties. For example, the AGNs reside mostly  in  massive galaxies\citep{Kauffmann03} and the strength of the nucleus activity is also correlated with its host property\citep{Kauffmann09}. For our sample of AGNs hosted by disk galaxies, we find that they are biased towards these hosts with large bulge component, high concentration and old stellar population.
To quantify the sample properties of these AGN hosts, we build a control sample of galaxies (without AGN phenomena) from the same  galaxy catalogue of SDSS, which  are selected to have the same sample size  and  the same distributions of stellar mass, redshift, size, concentration, $fracDev$ and $D_n(4000)$(the 4000\AA\, break spectra index, a rough stellar age indicator) as the AGN hosts. We believe that this control sample of non-active galaxies shares the same stellar properties and selection effects(e.g. redshift distribution) as the AGN hosts so that they would have the same $b/a$ distribution if their viewing angles were both random.

\begin{figure}
\includegraphics[width=80mm]{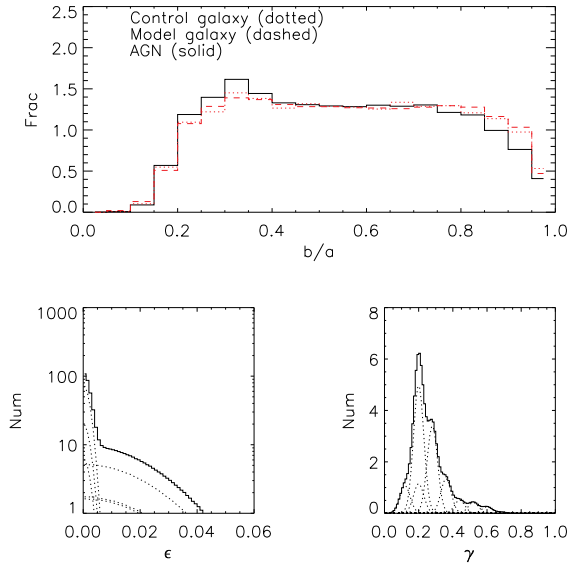}
\caption{The $b/a$, $\gamma$ and $\epsilon$ distributions of the AGN hosts and control galaxy sample. Top panel:  the $b/a$ histograms  of the AGN hosts(solid) and  control galaxies(dotted).  The dashed line shows the fitted histogram of the control galaxy sample from the non-negative least square regression. Bottom: The  $\gamma$ and $\epsilon$ distributions reproduced from the non-negative linear regression, where the dotted lines show the contributions from different basis components(see Section 3).}
\label{btoa}
\end{figure}

We compare the $b/a$ distributions of the AGN hosts to the control non-active galaxies. The result is shown in the top panel of Fig. \ref{btoa}. The AGN sample(solid histogram) has systematically more numbers of galaxies in smaller $b/a$ bins than the control galaxy sample(dotted histogram).  For a  better view of this over-density, we show the number ratios of the AGNs to the control galaxies ($N_{AGN}/N_{GAL}$) in different $b/a$ bins in Fig. \ref{model}. As one can see,  the $N_{AGN}/N_{GAL}$ decreases with the increasing of $b/a$ systemtically. The type 2 AGNs show both an excess of edge on objects and deficit of round objects. Since the galaxy sample have already been controlled to have the same physical properties and selection effects as the AGN hosts, the systemical bias in $b/a$ distribution of the AGN hosts could only be stemmed from their preferred (non-random) viewing angles. The bias towards the small axis ratios for the AGNs  is basically consistent with the scenario that the nuclear obscurer is partly co-aligned with the stellar disk.

\section{the shape of the  AGN hosts}

To quantify the preferred inclinations of the type 2 AGN hosts, the distributions of their shape parameters($\gamma$,$\epsilon$) need to be known in advance.

For the disk galaxies in general, it is found that their $\gamma$ distribution can be approximated by a normal function whereas  $\epsilon$ follows a normal or log-normal distribution\citep{Lambas92,Ryden04}. However, for our control sample of galaxies, the sample selection criteria is not well-defined and the goodness of fit is also bad when only single Gaussian distributions are assumed for $\gamma$ and $\epsilon/\rm{ln}\, \epsilon$. As an alternative, the control galaxy sample can be viewed as a combination of sub-samples of galaxies, where the $\gamma$ and $\epsilon$ distributions both follow Gaussian  distributions for each sub-sample. The $\gamma$ and $\epsilon$ distributions of the control galaxy sample are then  the sum of the Gaussian distributions of these sub-samples.

We first assume 10 basis distributions for $\gamma$ and $\epsilon$ each and  then  get 100 combinations of $(\gamma, \epsilon)$ distributions.  Specifically, for ten $\gamma$ distributions, the dispersions are fixed to 0.02, while the mean values are assigned from 0.1 to 0.5 with a step of 0.04. For $\epsilon$, the peaks of the Gaussian distribution are fixed to 0 and the dispersions are taken values from 0.02 to 0.2 with a step of 0.02. Each of the $(\gamma, \epsilon)$  combination is then viewed as a sub-sample. Using Monte-Carlo simulations of random viewing angles, we obtain the $b/a$ distributions of each sub-sample. Finally, we use the non-negative least square linear regression technic to fit the observed $b/a$ distribution of  the control galaxies to get the fraction of each sub-sample. The final fitted $b/a$ is shown as the dashed line in Fig. \ref{btoa}. As we can see, the fitting is very well, which is consistent with the observed value inside Possion error in most of the $b/a$ bins. The resulted $(\gamma, \epsilon)$ distributions of the control galaxy sample are shown in the bottom two panels of Fig. \ref{btoa}, where the dotted lines in each panel show the contributions from  different sub-samples.

With  $(\gamma, \epsilon)$ distributions known for the control galaxies, so that also for the AGN hosts, we move to model the viewing angles of AGN hosts.

\section{the orientation of the nuclear obscurer}

The biased axis ratio distribution (towards small values) of the AGN hosts implies that either the orientation of the molecular torus is correlated with the stellar  disk or there is an extra dust layer aligned with stellar disk. We denote these two different obscuring scenarios as the `aligned torus model' and `stellar dust model' respectively below.

We parameterize  the inclination of the dust torus as $i_T$, where $i_T$ ($0^\circ\le i_T\le 90^\circ$) is the angle between the normal of the  torus plane and the line-of-sight. Correspondingly, the inclination of the stellar disk is denoted as $i_S$. The tilt angle between the torus plane and stellar disk is then denoted as $\Delta_{TS}$.

\subsection{the obscurer: aligned torus model}\label{torus}

\begin{figure}
\includegraphics[width=80mm]{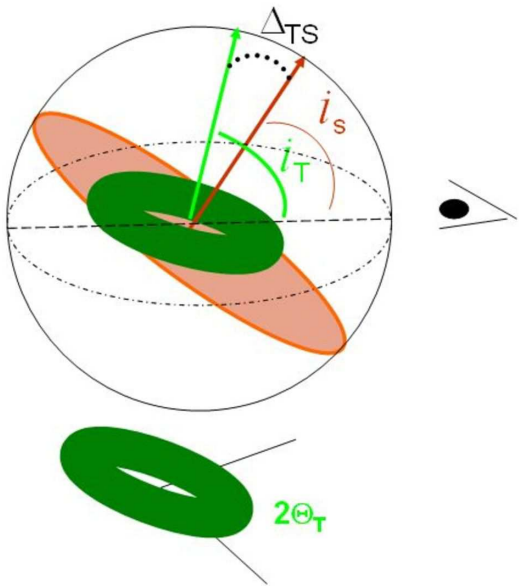}
\caption{The cartoon of the tilt angle ($\Delta_{TS}$) between the torus plane (inclination: $i_T$) and stellar disk (inclination: $i_S$) of a type 2 AGN ($90^\circ-i_T < \Theta_T$). }
\label{cartoon}
\end{figure}

In the aligned torus model, we assume that the obscuring torus has an opening angle $2\Theta_T$.
Obviously, a type 2 AGN satisfies the criteria $90^\circ-i_T < \Theta_T$. A cartoon of  such a torus model is shown in Fig. \ref{cartoon}.
The $\Theta_T$ and the tilt angle $\Delta_{TS}$ are then the two free parameters in this model.

$\Theta_T$ could be constrained from the type 2 AGN fraction $f_2$ since $f_2=sin(\Theta_T)$. The studies on the local Seyfert galaxies and the statistics of active galaxies in SDSS have shown that  $f_2$ is in the range of 70--90 percent for the low luminosity AGNs \citep{Ho97,Simpson05} , corresponding to  $45^\circ <\Theta_T<65^\circ$. Thus, we take  $45^\circ$ and $65^\circ$ as the low and up limits of  $\Theta_T$. Inside this range, $\Theta_T$ is set as  a free parameter.

For tilt angle $\Delta_{TS}$, it is a reasonable assumption that the most probable orientation of the torus plane is to be the same as the stellar disk, i.e. $P_{max}(i_S)=P(\Delta_{TS}=0)$ , but with significant scatters on $\Delta_{TS}$. On the other hand, $cos(\Delta_{TS})$ would follow a uniform distribution between 0 and 1 if the orientations of the torus plane and the stellar disk were independent. Thus, we further assume that $cos(\Delta_{TS}$) follows a Gaussian distribution which peaks at $\Delta_{TS}=0$ and has a scatter $cos(\Delta_{TS,m})$. Under this assumption, $\Delta_{TS,m}$ could be viewed as an effective tilt angle, within which 68 percent of the $\Delta_{TS}$ are smaller than.

With $\Theta_T$ and $\Delta_{TS,m}$ assigned, we make Monte-Carlo simulations to predict the number ratios $N_{GAL}/N_{GAL}$ in different $b/a$ bins.
In specific, we first generate a large Monte-Carlo sample of galaxies with random viewing angles and $(\gamma, \epsilon)$ distributions as shown in Fig. \ref{btoa}. For each galaxy with viewing angle $(i_{S,i},\phi_{S,i})$, we get the inclination of the torus plane $i_{T,i}$ from the tilt angle $\Delta_{TS,i}$ by assuming a random $\phi_{T,i}$, where $\phi_{T,i}$ and $\phi_{S,i}$ are the position angles of the torus plane and stellar disk respectively. After that, we get a modelling type 2 AGN sample by applying the criteria $90^\circ-i_{T,i} < \Theta_T$ for all Monte-Carlo galaxies. The other galaxies with $90^\circ-i_{T,i} > \Theta_T$ are classified as type 1 AGNs.

In the observed type 2 AGN sample, there are contaminations of intermediate type AGNs, which show weak broad wings on the H$\alpha$ lines but are classified as `galaxies' by the SDSS spectroscopic pipeline.  The fraction of these intermediate type contaminations is estimated to be about 8 percent from eyeball classifications\citep{Kauffmann03,Simpson05}.
To mimic this intermediate type AGN contaminations in our model as well, we also put some type 1 AGNs into our type 2 AGN sample in random as contaminations,  up to the maximum of 8 percent of the total sample.

\begin{figure}
\includegraphics[width=80mm]{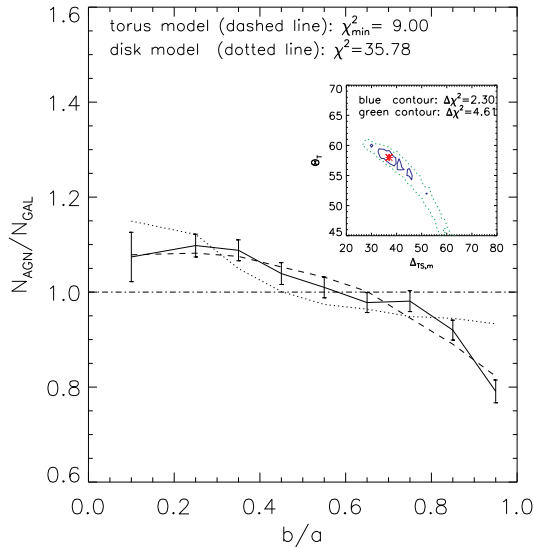}
\caption{The ratio of the numbers of the AGNs to the control galaxies in different $b/a$ bins. The solid line with error-bars shows the observational results from the SDSS. The dashed curve shows the best model fitting from the aligned torus model(see Section \ref{torus}) while the dotted curve show the model prediction from the stellar dust model(see Section \ref{stellar}). The dot-dashed horizontal line is constant 1. The sub-panel shows the confidence level of the two model parameters, $\Theta_T$ and $\Delta_{TS,m}$ for the aligned torus model.
}
\label{model}
\end{figure}

For each  Monte-Carlo  simulation, we randomly select 50,000 objects from the resulted library of type 2 AGNs(including type 1 contamination).
For each parameter set of $\Theta_T$ and $\Delta_{TS,m}$, we run the simulation 30 times to account for the Possion fluctuation.  We normalize the $b/a$ histograms of the model galaxies to be the same as the model AGNs and  calculate  $N_{AGN}/N_{GAL}$ in $b/a$ bins as that done for observations. The mean values of the 30 simulations are taken as the model output for each parameter set. Finally, we use the least square routine to search the best model parameters of $\Theta_T$ and $\Delta_{TS,m}$ by fitting the observed $N_{AGN}/N_{GAL}$ in different $b/a$ bins. The best model fitting is shown as the dashed line in Fig. \ref{model}, which has
the minimum $\chi^2_{min}\approx 9$,  comparable to the number of data points(9 $b/a$ bins), indicating an excellent goodness of fit.

The best model parameters are $\Theta_T=58^\circ$ and $\Delta_{TS,m}=37^\circ$ respectively. $\Theta_T=58^\circ$ corresponds to a type 2 AGN fraction of $f_2=0.85$, in excellent agreement with the results  quoted in literature for the low luminosity AGNs\citep{Simpson05,Lu10}. For the tilt angle between stellar disk and torus plane, the effective tilt angle  $\Delta_{TS,m}=37^\circ$ corresponds to a mean tilt angle $\bar{\Delta}_{TS}\approx 30^\circ$. We emphasis here that our assumption of the Gaussian distribution of $cos\Delta_{TS}$ is in a arbitrary way, however, the mean tilt angle ($\bar{\Delta}_{TS}\sim 30^\circ$) is quite robust and independent of the shape of the $\Delta_{TS}$ distribution.

 We show the 68 ($\Delta \chi^2=2.30$) and 90 ($\Delta \chi^2=4.61$) percent of the confidence levels of our model parameters in the sub-panel of Fig. \ref{model}. As we can see, the parameter $\Theta_T$ is  degenerated with $\Delta_{TS,m}$ to a certain extent.  The effect of increasing the torus opening angle  $\Theta_T$ is similar to the increasing of the effective tilt angle $\Delta_{TS,m}$, which both result in a more randomized  viewing angles of the disks of AGN hosts. Therefore, an independent measurement of $\Theta_T$ from the type 2 AGN fraction is crucial to constrain the tilt angle $\Delta_{TS}$ in our model.

\subsection{the obscurer: stellar dust model}\label{stellar}

In this section, we test the scenario that the over-density of the edge-on galaxies of type 2 AGNs is caused by an extra dust layer aligned with the stellar disk. Therefore, we no more assume that the nuclear obscurer is  aligned with the stellar disk as that in the aligned torus model(Section \ref{torus}), but consider  it being randomly oriented. Besides the random orientated torus,  the dust  aligned with the stellar disk will  obscure the nucleus when it is viewed as edge-on.

Using the denotations as in aligned torus model, a randomly oriented nuclear torus means that the orientations of the stellar disk and torus plane are independent of each other, i.e. $cos(\Delta_{TS})$ follows a random distribution between 0 and 1. For its opening angle, we set $\Theta_T=45^\circ$,  the lower limit of the aligned torus model, since we have an extra component of type 2 AGNs obscured by the host-aligned dust layer. For host-aligned obscurer, we assume its height is the same as the stellar  disk. Thus, for a stellar disk with height $\gamma$, its nucleus will be obscured when its viewing angles satisfies tan($i_S$)$<\gamma$.

We then combine these two types of obscurers and get the final type 2 AGN sample. The other model settings and procedures  are the same as in aligned tours model. We show the model prediction as the dashed line in Fig. \ref{model}. As we can see, this extra stellar obscurer model indeed predicts over-density of the low axis ratio galaxies for the AGN sample. However, this over-density mostly takes place at $b/a<0.4$. For the others with $b/a>0.4$, the predicted $N_{AGN}/N_{GAL}$ is almost a constant, whereas the observations show continuous decreasing of type 2 AGNs in more round galaxies. This is because that most of the stellar disks having $\gamma<0.4$(see Fig. \ref{btoa}).  Therefore, this model prediction shows significantly poorer fitting($\chi^2\approx 36$, outside the 99.9\% confidence interval of the $\chi^2$ distribution with 8 freedoms) than the aligned torus model.

However, we would mention that there is no tuneable  parameter in this stellar dust model, whereas the aligned torus model has two free parameters. Setting $\Theta_T$ to be a free parameter in the stellar dust model does not help, because the torus was assumed to be orientated randomly. However, assuming that the stellar dust layer following the same shape of the stars might be too simplistic. A more sophisticated treatment of the distribution of the stellar dust might improve the fit(e.g. \cite{Maiolino95}), but is out of the range of this study.

\section{conclusion and discussion}

In this study, we select a sample of type 2 AGNs hosted by spiral galaxies from the DR7 of the SDSS. We build a control sample of non-active galaxies to the AGN sample by matching their observational and physical properties in detail.  By comparing with the control sample, we find that the AGN hosts are systematically biased to edge-on views.
By modelling  the height and ellipticity of the disk galaxies  with a non-parametric technic, we further find that this systematical bias  can not be quantified by an extra dust layer aligned with stellar disk but can be nicely reproduced by assuming an average tilt angle of $\sim 30$ degree between the torus plane and stellar disk.

Some of the assumptions in our model are simplified, e.g. the constant type 2 AGN fraction. There are studies suggesting that the opening angle of the torus is correlated with the central AGN luminosity. For high luminosity AGNs, the distance of the torus from the nucleus is further (`receding torus model') so that the torus opening angel  and the type 2 AGN fraction $f_2$ are smaller\citep{Lawrence91}. Most of our AGNs are low luminosity objects($L_{\rm{[OIII]}}< 10^8L_\odot$), and the type 2 AGN fraction has been shown to be roughly a constant in this luminosity region\citep{Simpson05,Lu10}. A separation of our sample into two sub-samples according to $L_{\rm{[OIII]}}$  also would not show any significant differences but reduce the goodness of the statistics.

In the unified AGN diagram, the torus lies on the same plane of the accretion disk. In this case, the mean tilt angle between the torus plane and the stellar disk constrained from our model could also be understood as the mean tilt between the accretion disk and stellar disk.

\section*{Acknowledgments}
We thank the referee for useful comments which significantly clarified the text.  S.S thanks Guniveror Kauffmann, Chenggang Shu and Xinwu Cao for helpful discussions. This project is supported by NSFC10803016, 10833005, 10703009, 10821302, 10833002, 10973028; \\NKBRSF2007CB815402, 2009CB824800, Shanghai Rising-Star Program(08QA14077) and Shanghai Municipal Science and Technology Commission No. 04dz\_05905.

\end{document}